\documentclass[aps,prl,twocolumn,superscriptaddress]{revtex4}

\usepackage{bm}
\usepackage{graphicx}
\graphicspath{{Figures/}} 
\usepackage{xcolor}
\usepackage{dcolumn}

\begin{document}
                                      
\title{An optical investigation of the heavy fermion normal state in superconducting UTe$_2$}

\author{Sirak M. Mekonen}
\affiliation{The Institute for Quantum Matter, Department of Physics and Astronomy, The Johns Hopkins University, Baltimore, MD, 21218, USA}

\author{Chang-Jong Kang} 
\affiliation{Department of Physics and Astronomy, Rutgers University, Piscataway, NJ 08856, USA}
\affiliation{Department of Physics, Chungnam National University, Daejeon, 34134, South Korea}

\author{Dipanjan Chaudhuri}
\affiliation{The Institute for Quantum Matter, Department of Physics and Astronomy, The Johns Hopkins University, Baltimore, MD, 21218, USA}

\author{David Barbalas}
\affiliation{The Institute for Quantum Matter, Department of Physics and Astronomy, The Johns Hopkins University, Baltimore, MD, 21218, USA}

\author{Sheng Ran}
\affiliation{Center for Nanophysics and Advanced Materials, Department of Physics, University of Maryland, College Park, MD, 20742, USA}

\affiliation{NIST Center for Neutro Research, National Institute of Standards and Technology, Gaithersburg, MD, 20742, USA.}

\affiliation{Department of Materials Science and Engineering, University of Maryland, College Park, MD, 20742, USA.}

\author{Gabriel Kotliar}
\affiliation{Department of Physics and Astronomy, Rutgers University, Piscataway, NJ, 08856, USA}

\affiliation{Department of Condensed Matter Physics and Materials Science,Brookhaven National Laboratory, Upton, NY, 11973, USA}

\author{Nicholas P. Butch}
\affiliation{Center for Nanophysics and Advanced Materials, Department of Physics, University of Maryland, College Park, MD, 20742, USA}

\affiliation{NIST Center for Neutro Research, National Institute of Standards and Technology, Gaithersburg, MD, 20899, USA.}

\affiliation{Department of Materials Science and Engineering, University of Maryland, College Park, MD, 20742, USA.}

\author{N.P. Armitage}
\affiliation{The Institute for Quantum Matter, Department of Physics and Astronomy, The Johns Hopkins University, Baltimore, MD, 21218, USA}


\date{\today}

\begin{abstract}
The recently discovered superconductor, UTe$_2$, has attracted immense scientific interest due to observations that suggest odd-parity superconductivity. It is believed that the material is a heavy-fermion metal at low temperatures although details of the normal state are unclear.  Using Fourier transform infrared spectroscopy (FTIR), we investigated the  normal state electronic structure of UTe$_2$ at zero applied magnetic field. Combining the measured reflectivity with the dc resistivity, the complex optical conductivity was obtained over a large frequency range. The frequency dependence of the real part of the optical conductivity exhibits a MIR peak around 4000 cm$^{-1}$ and a narrow Drude peak that develops below 40 K. A combination of density functional and dynamic mean field theory (DFT + DMFT) gives spectra in close correspondence to the experiment. Via this comparison we attribute the prominent MIR peak to inter-band transitions involving a narrow U 5$f$ feature that develops near the Fermi level. In this regard, this comparison along with data that shows the scale of the low temperature mass renormalization gives spectroscopic evidence for the existence of a low energy Kondo resonance at temperatures just above the onset of superconductivity and implicates heavy electrons in the formation of the superconducting state.  We find that the coherent Kondo resonance is primarily associated with a collapse of scattering and less with a transfer of spectral weight.

\end{abstract}

\maketitle


\maketitle



Heavy fermion compounds are comprised of a lattice of strongly correlated \textit{f}-electrons coupled to a sea of conduction electrons. At high temperatures, these systems behave as an ensemble of weakly interacting fermions with modest masses and localized \textit{f}-electrons. At low temperatures, Kondo hybridization between these subsystems can give a metallic state with charge carriers whose masses can be a thousand times the free electron mass~\cite{Stewart1984}. The high temperature state is reminiscent of the situation for isolated magnetic impurities, where the Kondo interaction causes spin scattering of conduction electrons. However, at low temperatures, this interaction causes the conduction electrons to collectively and coherently screen the local magnetic moments of the \textit{f}-electrons of the lattice to form a Kondo singlet. This effect leads to the formation of narrow resonance at the Fermi level, which is often referred to as the Abrikosov-Suhl or Kondo resonance~\cite{Stewart1984,degiorgi1999,scheffler2013, bosse2012}. Such systems frequently host other novel states of matter including various magnetic orders and unconventional superconductivity~\cite{Stewart1984}.

Uranium-based $5f$ heavy fermion compounds such as UPt$_3$, UGe$_2$, URhGe, and UCoGe~\cite{Stewart1984,Saxena2000,Aoki2001,Huy2007} have been found to host unconventional superconductivity. The coexistence of (or proximity to) ferromagnetism makes them strong candidates to realize odd-parity superconductivity~\cite{Jiao2019}. Odd-parity superconductors are of fundamental interest for use in quantum computation due to their capacity to host topologically protected excitations~\cite{Sato2017, Beenakker2013, Sarma2015, Alicea2012}. UTe$_2$ with a transition temperature (T$_{sc}$) of 1.6 K, has attracted immense scientific interest due to observations that suggests odd-parity pairing: its $b$ direction critical field (H$_{c2}$) exceeds the Pauli limit by more than an order of magnitude~\cite{ran2019nearly, Aoki2019}; two re-entrant SC phases are observed in high magnetic field~\cite{ran2019extreme}; there are strong nearly critical ferromagnetic fluctuations~\cite{ran2019extreme, Tokunaga2019,Sundar2019}; the Knight shift is anomalously constant through the superconducting transition~\cite{ran2019nearly, Tokunaga2019, nakamine2019}. Additionally, thermal transport, heat capacity, and magnetic penetration depth measurements suggest point-like nodal structures~\cite{Metz2019}, scanning tunneling microscopy suggests unconventional Cooper pairing~\cite{Jiao2019} and microwave experiments suggest a large normal fluid response~\cite{Bae2019}. 

\begin{figure}[t]
   \includegraphics[width=0.9\columnwidth]{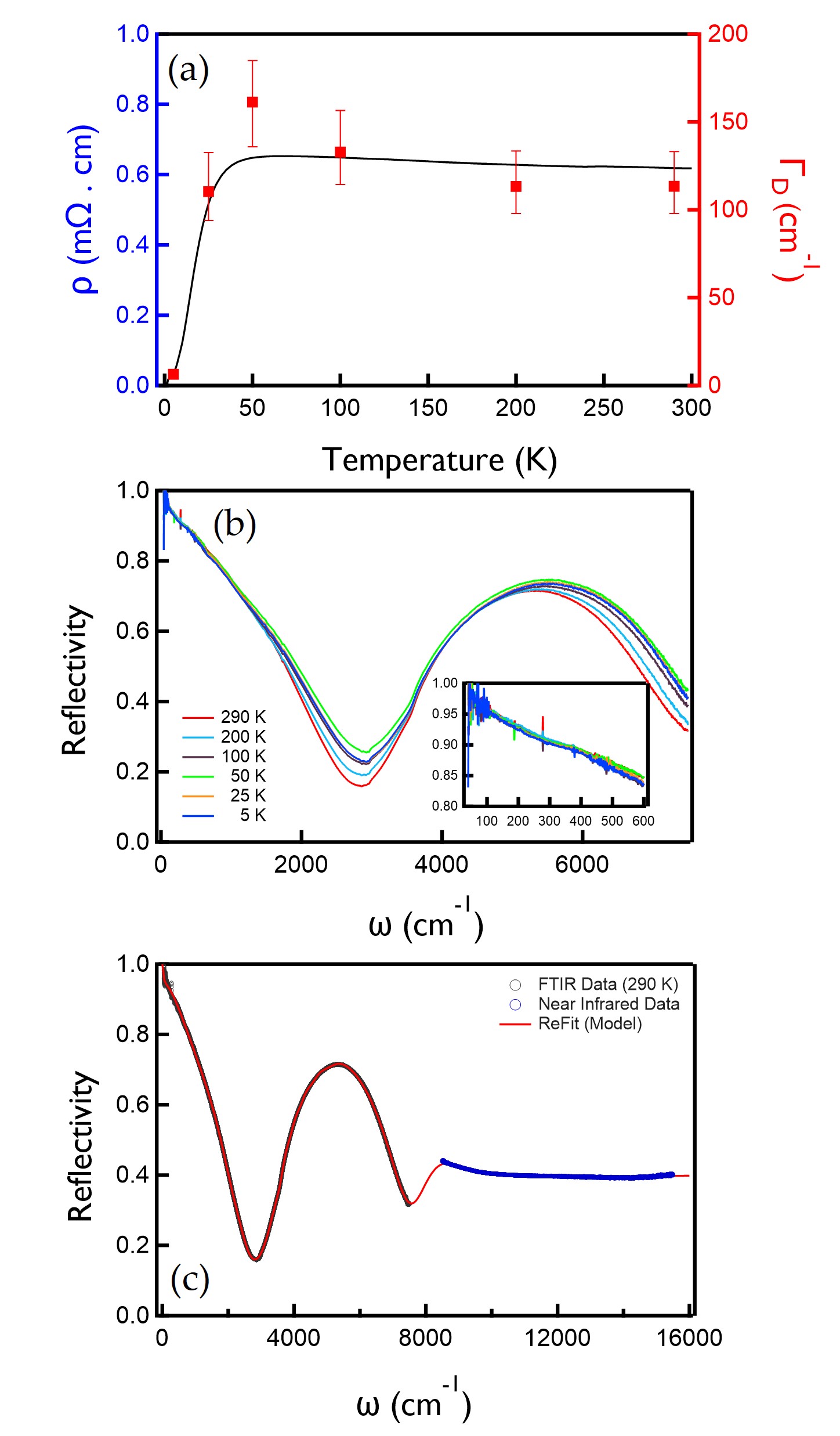} 
\caption{ (a) 4-probe dc resistivity of UTe$_2$ single crystals with (100) current, plotted alongside the extracted narrowest Drude scattering rate. (b) Reflectivity of UTe$_2$ for the investigated temperatures. The inset shows the low-frequency temperature dependent reflectivity measurements. (c) Reflectivity data at 290 K along with fit from RefFit.}
\label{fig1}
\end{figure}

In contrast to the extensive experiments in the superconducting state, there is less known about UTe$_2$'s normal state. Its resistivity is weakly increasing with decreasing temperature until approximately 40 K (Fig.\ref{fig1}(a)), where it then falls dramatically~\cite{Shlyk1999, ran2019nearly,eo2021}.
This is typical for many heavy fermion systems, where conduction electrons scatter incoherently from localized moments at higher temperature, but coherently hybridize with localized spins below a lattice coherence scale T$_{KL}$~\cite{Stewart1984,degiorgi1999}.  In UTe$_2$, a Curie-Weiss susceptibility is found above 150 K with an effective moment close to the 5\textit{f$_2$} or 5\textit{f$_3$} free-ion value and a Weiss constant that is -80 to -130K (depending on the direction). Despite the large Weiss constant, no magnetic order is found down to low temperature~\cite{Ikeda2006}. At low temperature the susceptibility is anisotropic~\cite{Ikeda2006} with either a 40 K maxima for $H || [010] $ or increasing behavior for $H || [100]$ or $[001]$. The maximum for $H || [010] $ at 40K can be taken as a signature of heavy fermion formation and a low temperature crossover to itinerant 5$f$ electrons.  A Sommerfeld coefficient of $\gamma \approx $100-120 mJ/mol$\cdot$K$^2$ is likely indicative of heavy fermions~\cite{Metz2019,Aoki2019}.

However, while the formation of a heavy fermion normal state seems reasonable, details of the electronic structure remains unclear. UTe$_2$ crystallizes in the orthorhombic, centrosymmetric structure (space group 71 \textit{Immm}). The U atoms compose chains oriented along the [100] \textit{a} axis. The Te(2) atoms form a chain in the [010] direction. These quasi-1D structures are expected to be reflected in the electronic structure. The most straightforward band calculations predict that UTe$_2$ should be a narrow gap semiconductor~\cite{Aoki2019}. To understand the actual electronic structure of UTe$_2$, it is illuminating to calculate the bands for ThTe$_2$~\cite{harima2020obtain} as thorium favors a unfilled quadrivalent thorium 5$f_0$ state and has no 5$f$ electrons near the Fermi level. ThTe$_2$ is predicted to be a metal with conduction bands of Th 6$d$ and Te 5$p$ character. Therefore the predicted insulating character of UTe$_2$ originates in mixing between the conduction and U 5$f$ electrons~\cite{harima2020obtain}. UTe$_2$ can be made metallic in such calculations by adding a local Coulomb $U$ term for the 5$f$ electrons or by ad hoc shifting the 5$f$ levels by 0.1 - 0.2 Ry~\cite{harima2020obtain}. In either case, it indicates that band calculations treat the 5$f$ electrons improperly. Soft X-ray photoemission measurements suggested that U 5$f$ states of UTe$_2$ have an itinerant but strongly-correlated nature with enhanced hybridization of the Te 5$p$ states~\cite{Fujimori2019}. A study with high resolution ARPES reveal two light quasi-1d bands at the Fermi level, which are attributed to U and Te(2) chains. This study found good correspondence with density functional calculations combined with DFT + DMFT calculations~\cite{Miao2020}. However, signatures of the 5$f$ bands at E$_F$ were only tentatively attributed, and the related feature had little visibility.  Such experiments may be hampered by poor surface quality that is inherent to cleaving reasonably isotropic materials. Thus, there is a need for probes with bulk sensitivity.

We have measured optical reflectance on the (100) shiny as-grown surface ($\approx2$ mm $\times$ $2$ mm $\times$ $2$ mm) of a high quality UTe$_2$ crystal using FTIR spectroscopy with unpolarized light. The reflectivity of single crystal UTe$_2$ was measured using a commercial FTIR spectrometer (Bruker, Vertex 80V, Source: Hg Arc Lamp/Globar, Detectors: Bolometer/DLaTGS/MCT) across the far and mid infrared spectral ranges spanning from 40 cm$^{-1}$ to 16,000 cm$^{-1}$ (i.e. $1.2$-$480$ THz) for temperatures between $5$ K to $300$ K. To extend the measurement across a broader spectral range, the infrared spectra was supplemented by near infrared spectrum between 8500 cm$^{-1}$ and 16000 cm$^{-1}$ measured at room temperature, using the Bruker VIS I spectral range extension. The reflected signals at each temperature were referenced to a gold film that was deposited on the sample \textit{in situ} and then corrected by the known reflection coefficient of gold over the full spectral range\cite{Chaudhuri2017}. In addition to the optical measurements, conventional 4-probe dc resistivity was measured. Single crystal UTe$_2$ was grown by chemical transport using iodine as a transport agent.  For further details see Ref.~\cite{Hutanu2019}.

In Fig. \ref{fig1}(b) we show the measured reflectivity from 40 to 7500 cm$^{-1}$. Consistent with the metallic nature of these systems, the reflectivity approaches unity at low frequencies and has a sharp plasma edge-like feature around $2500$ cm$^{-1}$. With decreasing temperature, the reflectivity first increases for all frequencies until 50K, and then decreases. A prominent broad peak with a slightly temperature dependent maximum is found in the reflectivity $\approx$ 5500 cm$^{-1}$. These subtle temperature dependent changes can be contrasted to the dc resistivity that shows a dramatic drop below 40K.  We extracted the complex optical conductivity by fitting the reflectivity and dc resistivity simultaneously to a complex multi-oscillator model

\begin{equation}
    \centering
   \sigma(\omega) = i \epsilon_0 \omega \Big[\sum_i \frac{\omega_{pi}^2}{\omega^2 - \omega_{0i}^2 + i \omega \Gamma_i } -  \epsilon_\infty + 1  \Big],
\end{equation}
where $\omega_{pi}$ is the plasma frequency of the $i$th component, $\omega_{0i}$ is its oscillation frequency, and $\Gamma_{i}$ is its scattering rate. This is particularly useful procedure in dealing with data sets where different quantities are obtained in  different frequency ranges. Here we fit with nine finite frequency Lorentzian oscillators and two zero frequency "Drude" (e.g. $\omega_{0i} = 0$) oscillators, a narrow one and a wide one. The primary purpose of this fitting and parameterization is the extraction of the \textit{complex} conductivity in a Kramers-Kronig consistent fashion and except for the Drude components, the particular values of the fit parameters do not necessarily have any physical significance.  Any Kramers-Kronig consistent set of complex functional forms would suffice to generate the conductivity. In Fig. \ref{fig1}(c) we show the 290 K reflectivity along with the output of complex fitting. The fits to the reflectivity (constrained by the dc measurement) is excellent over the entire frequency range.

\begin{figure}[t]
    \includegraphics[width=0.9\columnwidth]{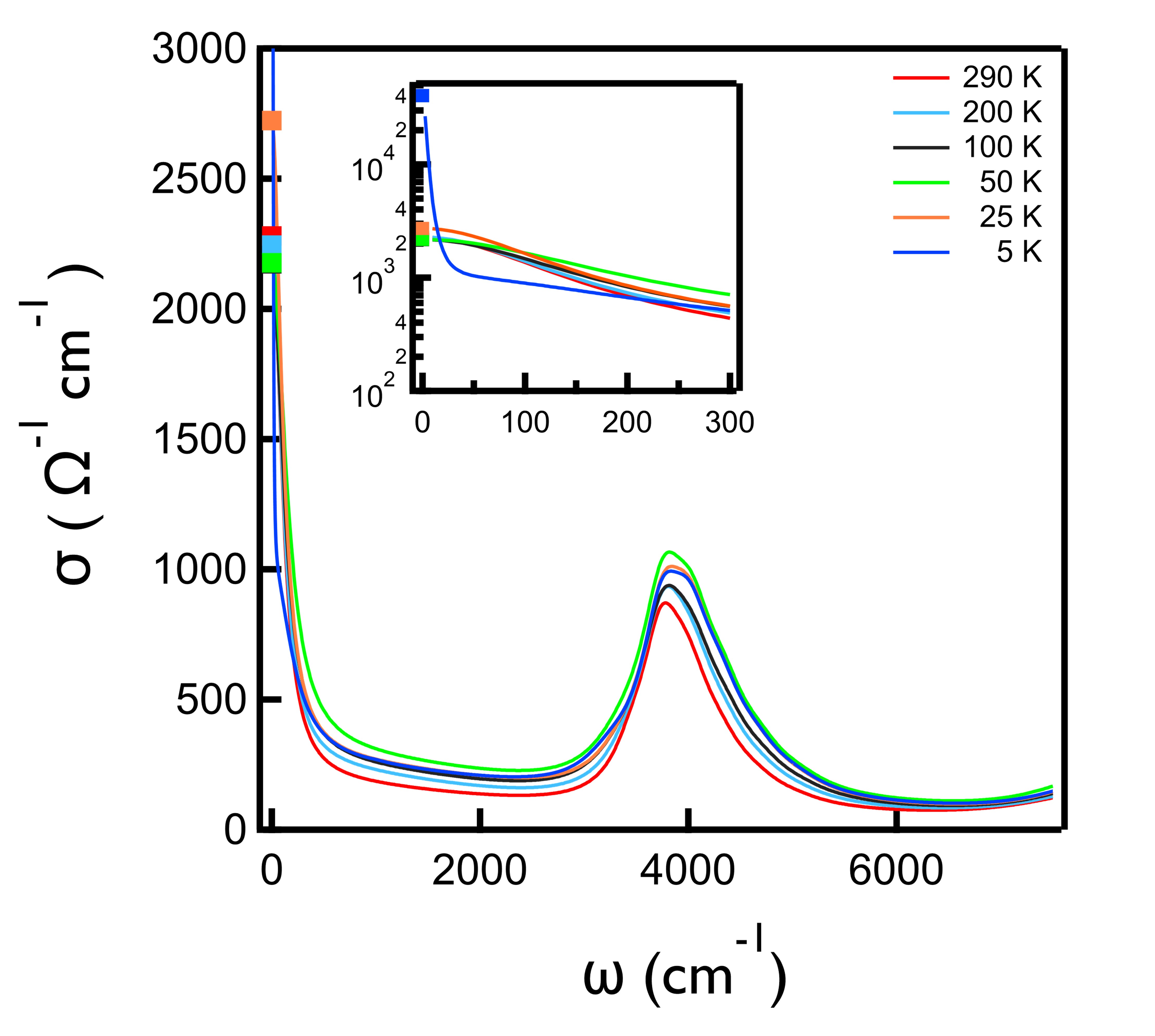}
    \caption{Real part of conductivity. The square  markers represent the dc conductivity ($\omega$ = 0) calculated from Fig. \ref{fig1}(a). In the inset the dc point for 5K is shown and has a value of 40500 $\Omega^{-1}$cm$^{-1}$.   It is too a high value to be displayed on the linear scale plot.}
    \label{fig2}
\end{figure}

In Fig. \ref{fig2}, we show the extracted real optical conductivity from the simultaneous complex fitting of the reflectivity and dc resistivity for different temperatures. The conductivity is characterized by a strongly temperature dependent zero-frequency Drude-like peak and a weakly temperature dependent maximum at $\approx$ 4000 cm$^{-1}$ (0.49 eV). As temperature decreases from 290 K, the maximum first gains spectral weight and shift to slightly higher frequency, then looses spectral weight below 40 K.

\begin{figure}[ht]
    \includegraphics[width=0.9\columnwidth]{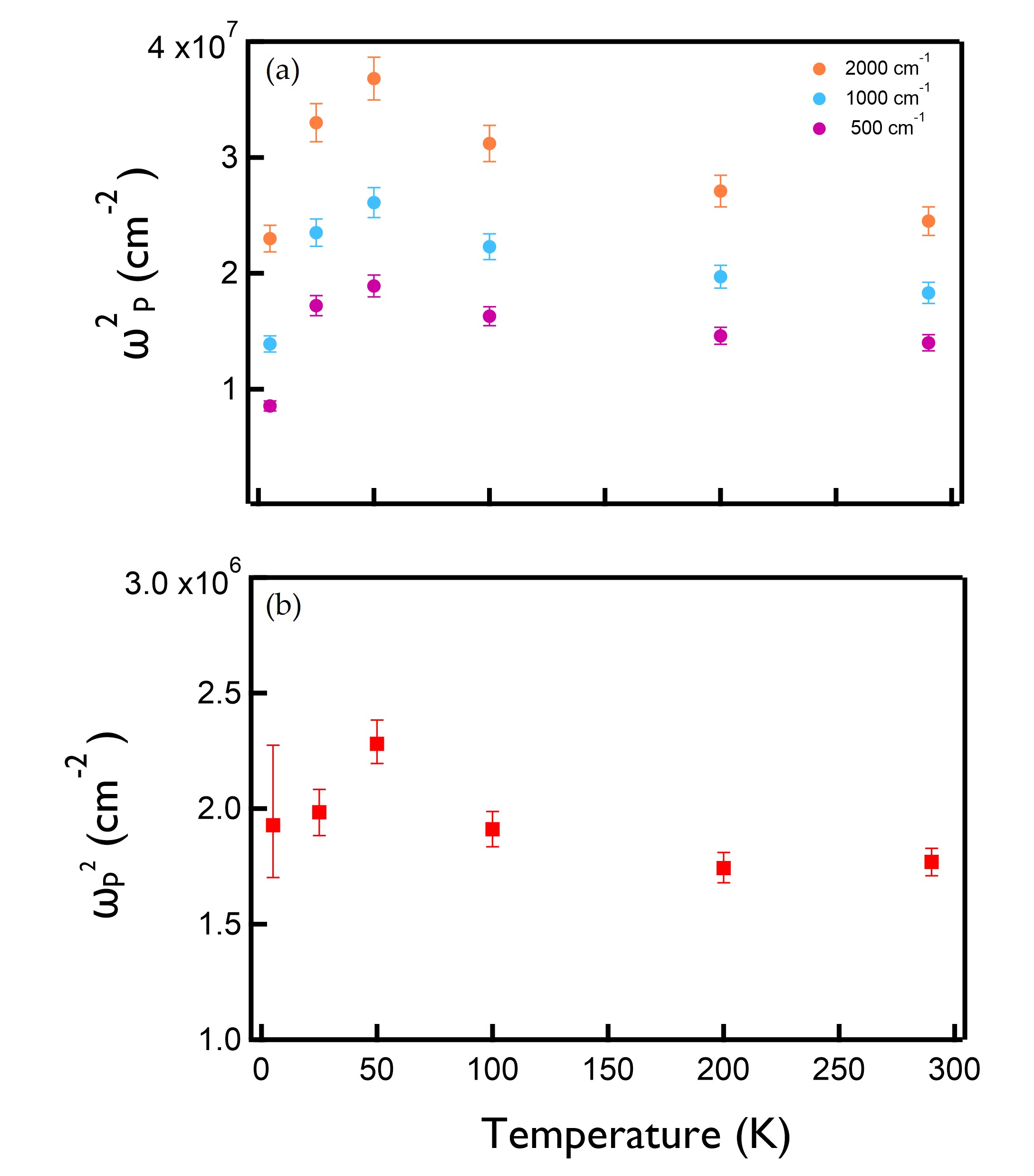}
    \caption{(a) Integrated spectral weight (in units of the plasma frequency squared) as a function of temperature using different cut-off frequencies. (b) Spectral weight of narrow Drude peak.}
    \label{fig3}
\end{figure}

\begin{figure}[ht]
    \includegraphics[width=0.9 \columnwidth]{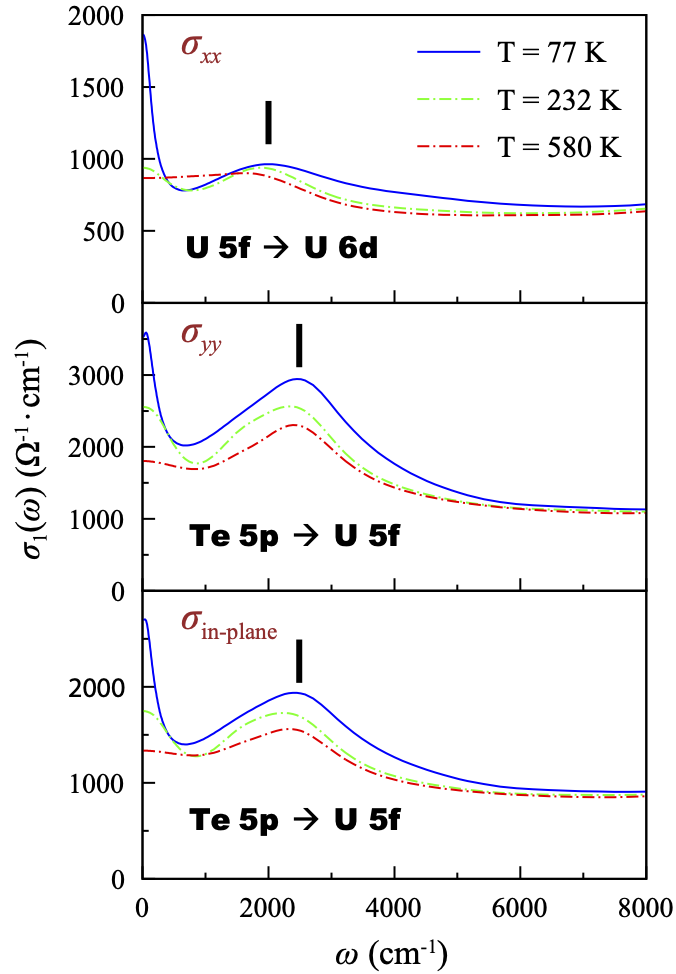}
    \caption{Optical conductivities along $xx$, $yy$, and in-plane (averaged) polarizations computed within DFT+DMFT at different temperatures. The solid black line indicates the peak position of the interband optical transition.}
    \label{dmft-optics}
\end{figure}

The prominent zero frequency peak sharpens dramatically at low temperatures (Fig. \ref{fig2} inset). We captured its dependencies via plotting the Drude parameters of the narrowest Drude. The wider Drude had a width of $\approx$ 2400 cm$^{-1}$, which gives a weakly temperature dependent feature with little frequency dependence up to the range of the interband transitions. In contrast, the narrow Drude has a very temperature dependent width and spectral weight. We plot its width and spectral weight in Figs.~\ref{fig1}(a) and \ref{fig3}(b). At low temperatures the narrow Drude peak becomes extremely narrow and loses some spectral weight below 50 K. The former behavior can be attributed to the formation of heavy quasiparticles due to the formation of a 5$f$ Kondo resonance. Note that this behavior is uncommon for other heavy fermions where the heavy fermion state usually manifests as the emergence of a narrow peak out of a wider background \cite{bosse2012,degiorgi1999,donovan1997observation}. Here, the principal effect appears to be a dramatic sharpening of this peak.  The origin of the  spectral weight decrease below 50K is unclear, but could be associated with a precursor to a hybridization gap~\cite{dordevic2001hybridization}.

The ratio of the total low frequency intraband spectral weight $( \omega_{p \; Total}^2)$ (which represents the uncorrelated band) to the narrow Drude's spectral weight $( \omega_{p \; Narrow}^2)$ can be taken as a rough measure of the heavy fermion mass renormalization~\cite{bosse2012}. In Fig.~\ref{fig3}(a), we plot the total spectral weight (in units of cm$^{-2}$) as a function of temperature for different cut-off frequencies. We somewhat arbitrarily take the maximum cut-off frequency for the total intraband spectral weight to be 2000 cm$^{-1}$ as this is below the interband transitions, but well above the narrow Drude regime. Using the value for the maximum total intraband spectral weight and the low temperature narrow Drude spectral weight, we find $\omega_{p \; Total}^2 / \omega_{p \; Narrow}^2 = m^*/m_b \approx 25$, where $m_b$ is the band mass.   If one takes as measure of the uncorrelated spectral weight the computed angular averaged spectral weight of ThTe$_2$ is 1.23$\times10^9$cm$^{-2}$~\cite{Kim21a}, and computes the same average one finds a mass renormalization (again with respect to the band mass) of 65.  These estimates are of order the mass renormalizations found in the heat capacity~\cite{Metz2019,Aoki2019}, but note that they are lower bounds on the mass renormalization as there could still be the formation of even narrower Drude-like features with less spectral weight that would lead to larger masses. To resolve these, careful microwave studies of the normal state conductivity must be done.  Finally, note that the temperature dependence of the narrow Drude scattering rate,  $\Gamma_D$ scales as the resistivity with a notable 40 K drop (Fig. \ref{fig1}(a)). The similarity in temperature dependence of resistivity and $\Gamma_D$ points to the coherent nature of the propagating quasiparticles. 

\begin{figure*}
    \includegraphics[width=0.9\textwidth]{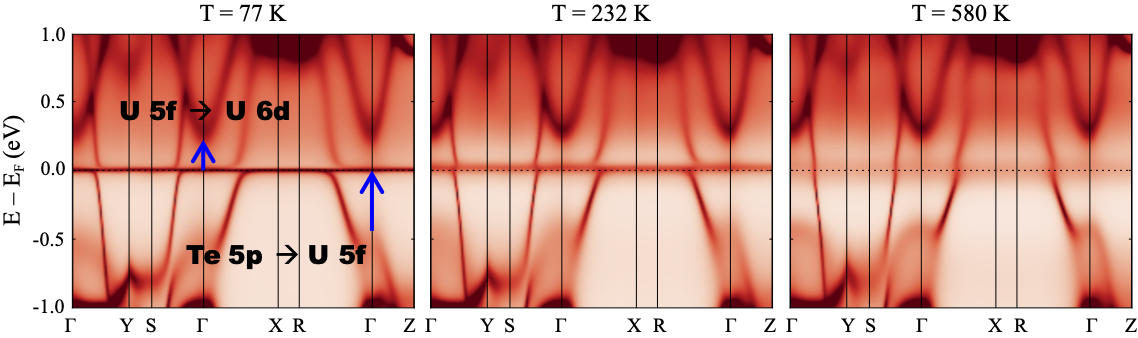}
    \caption{Momentum-resolved spectral functions $A(k,\omega)$ at several temperatures computed within DFT+DMFT. Two optical transitions are presented with blue arrows.}
    \label{dmft-band}
\end{figure*}

To get further insight into this data, we performed fully charge self-consistent DFT+DMFT calculations~\cite{Georges1996, Kotliar2006, Held2007,Haule2010} implemented in the Wien2k package~\cite{Blaha2020} with experimentally determined lattice. Continuous-time quantum Monte Carlo (CTQMC)~\cite{Werner2006, Haule2007} was implemented with a local impurity solver. We choose a wide hybridization energy window from -10 eV to 10 eV with respect to the Fermi level E$_{F}$. The fully rotationally invariant form was applied for a local Coulomb interaction Hamiltonian with on-site Coulomb repulsion U = 6 eV and Hund's coupling J$_{H}$ = 0.57 eV, where the U and J$_{H}$ values reproduce aspects of the recent ARPES data~\cite{Miao2020}. The maximum entropy method~\cite{Jarrell1996} was used for analytical continuation to obtain the self-energy on the real frequency axis. Unfortunately in multi-orbital systems the CTQMC develops increasing statistical noise in the calculation as temperature is lowered, which prevents accurate numerical results below the 77 K displayed.

The DFT+DMFT optical conductivity is computed as shown in Fig.~\ref{dmft-optics} (We show even higher frequencies in the Supplementary Materials).  Three different polarization configurations are presented. The optical conductivities $\sigma_{xx}$ and $\sigma_{yy}$ and their average are shown in panels Fig.~\ref{dmft-optics} a,b, and c respectively. $xx$ and $yy$ polarizations give peaks at $\approx$2000 and $\approx$2500 cm$^{-1}$ respectively. These peaks increase spectral weight and sharpen as temperature is lowered. In addition, a sharp Drude peak develops upon cooling, which is related to the formation of coherent U $5f$ bands near the Fermi level at low temperature (see Fig.~\ref{dmft-band}). There is a close correspondence with the experiment. The averaged in-plane optical conductivity, which is obtained from $(\sigma_{xx}+\sigma_{yy})/2$, is provided in Fig.~\ref{dmft-optics}(c) for comparison with the experiment. Since the optical spectral weight along $yy$ is larger than along $xx$, the inter-band optical peak of the in-plane optical conductivity mainly arises from the $yy$ polarization.  The origin of the two optical transitions could be assigned from inspection of the DFT+DMFT electronic structure in Fig.~\ref{dmft-band}. The optical peak realized with $xx$ polarization originates from U $5f$ to U $6d$ transition as shown, whereas the peak in the $yy$ channel arises from the Te $5p$ to U $5f$ transition. The polarization dichotomy arises from the presence of the U and Te chains that run along $x$ and $y$ directions, respectively. This is strong evidence that the Kondo hybridization yields coherent electronic bands 5$f$ band near the Fermi energy.  Note that the recent ARPES experiments did not observe the U $5f$ bands~\cite{Miao2020}, however the present optical conductivity experiment does. 

Despite close correspondence of the experiment with theory, there are still notable differences e.g. the numerical disagreement in the position of the MIR peak. Attempts to change model parameters down to $U=4$ eV, resulted in only modest increases of the peak position, but unacceptably large disagreements with the photoemission. Increasing the Coulomb repulsion to $U=8$ eV resulted in decreases of the MIR peak by approximately 0.2 eV, e.g. the wrong directions. It may be that LDA calculation are not sufficiently accurate to describe the underlying uncorrelated band structure and going to a `GW' method + DMFT would improve agreement. Alternatively, a better treatment of longer range interactions that are outside the DMFT could be important.

The presence of the sharp optical transition and the narrow Drude peak is evidence for formation of a 5$f$ derived Kondo resonance.  At low temperature and low energies, the narrow Drude becomes much sharper due to the formation of heavy fermion quasiparticles with contribution of $5f$ electrons. There is a small decrease in the total Drude spectral weight below 40K.  One other interesting and unexplained aspect to the data (both experiment and numerical) is the dichotomy between the relatively weaker temperature dependence of the MIR peak than the narrow Drude peak. This is presumably related to the fact that spectral weight of the narrow Drude peak has only a comparatively weak temperature dependence. Therefore the formation of the coherent state below 40K seems to be primarily associated with a collapse of the scattering rate and not a redistribution of spectral weight. This is different than many other heavy fermion compounds where there is a growth of very low frequency spectral feature with cooling. A comparison of the data to DFT + DMFT calculations shows that the MIR peak arises from interband transitions to or from the 5$f$ states. Our data gives spectroscopic evidence for the formation of a narrow low energy Kondo resonance at temperature just above the onset of superconductivity. This data is then the first spectroscopic observation that the superconductivity in this materials comes from well-formed 5$f$ derived heavy fermion quasiparticles.

Work at JHU was supported by NSF DMR-1905519 and an AGEP supplement. CJK and GK were supported by NSF  DMR-1733071. Work at UMD was supported by NIST. Identification of commercial equipment does not imply endorsement by NIST. We would like to thank S. Anlage, S. Bae, and A. Wray for helpful conversations.

\bibliography{UTe2NormalState}


\end{document}